%% file: simple.tex
\documentclass[twocolumn,10pt,aps,pra,superscriptaddress,nofootinbib]{revtex4-1}

\usepackage{scrextend}
\usepackage{extra_style}
\usepackage{longtable}
\usepackage[a4paper,bindingoffset=0.2in,left=.75in,right=.75in,
            top=1.5in,
            bottom=1.25in,
            footskip=.005in]{geometry}
\usepackage{graphicx}
\renewcommand{\thefootnote}{$\star$} 
\renewcommand*{\thefootnote}{\fnsymbol{footnote}}
\newcommand{\LLM}{AI-assisted}
\newcommand{\SuppI}{Appendix}

\usepackage{xspace}

\usepackage{silence}
\newcommand{\estimatedReviewsLLM}{15.8\%\xspace}

\newcommand{\estimatedEffectAcceptance}{3.1 percentage points\xspace}
\newcommand{\estimatedEffectAcceptancePval}{$p=0.024$\xspace}

\newcommand{\estimatedEffectAcceptanceOdds}{13.8\%\xspace}
\newcommand{\estimatedEffectAcceptanceOddsPVal}{$p=0.024$\xspace}

\newcommand{\estimatedEffectAcceptanceB}{4.9 percentage points\xspace}

\newcommand{\estimatedEffectAcceptanceBPVal}{$p=0.024$\xspace}
\newcommand{\estimatedEffectAcceptanceBOdds}{31.1\%\xspace}
\newcommand{\estimatedEffectAcceptanceBOddsPVal}{$p=0.031$\xspace}

\newcommand{\acceptanceRateB}{73.6\%\xspace}
\newcommand{\prevalenceB}{20.7\%\xspace}

\newcommand{\estimatedEffectScore}{14.4\%\xspace}
\newcommand{\estimatedEffectScorePVal}{$p=0.002$\xspace}
\newcommand{\estimatedEffectScoreIntepretable}{$53.4\%$\xspace}

\newcommand{\reviewstf}{$n=28{,}028$\xspace}
\newcommand{\submissionsAItfone}{$n=3{,}357$\xspace}
\newcommand{\submissionstf}{$n=7{,}404$\xspace}
\newcommand{\matchedSampletf}{$n=5{,}132$\xspace}
\newcommand{\fpratev}{1.7\%\xspace}
\newcommand{\MatMeth}{Materials and Methods\xspace}

\newcommand{\xhdr}[1]{\vspace{1.7mm}\noindent{{\bf #1.}}}

\begin{document}

\title{The AI Review Lottery: Widespread AI-Assisted Peer Reviews\\Boost Paper Scores and Acceptance Rates}
\author{ Giuseppe Russo Latona,$^*$ 
Manoel Horta Ribeiro,$^\dagger$
Tim R. Davidson,$^\dagger$\\
Veniamin Veselovsky,$^\dagger$
Robert West$^*$\\ 
{EPFL} \\
}

\begin{abstract}
\noindent 
Journals and conferences worry that peer reviews assisted by artificial intelligence (AI), in particular, large language models (LLMs), may negatively influence the validity and fairness of the peer-review system, a cornerstone of modern science.
In this work, we address this concern with a quasi-experimental study of the prevalence and impact of \LLM{} peer reviews in the context of the 2024 International Conference on Learning Representations (ICLR), a large and prestigious machine-learning conference.
Our contributions are threefold.
Firstly, we obtain a lower bound for the prevalence of \LLM{} reviews at ICLR 2024 using the GPTZero LLM detector, estimating that at least \estimatedReviewsLLM of reviews were written with AI assistance.
Secondly, we estimate the impact of \LLM{} reviews on submission scores. 
Considering pairs of reviews with different scores assigned to the same paper,
we find that in \estimatedEffectScoreIntepretable of pairs the \LLM{} review scores higher than the human review
(\estimatedEffectScorePVal; relative difference in probability of scoring higher:\ $+$\estimatedEffectScore in favor of \LLM{} reviews).
Thirdly, we assess the impact of receiving an \LLM{} peer review on submission acceptance. In a matched study, submissions near the acceptance threshold that received an \LLM{} peer review were \estimatedEffectAcceptanceB (\estimatedEffectAcceptanceBPVal) more likely to be accepted than submissions that did not.
Overall, we show that \LLM{} reviews are consequential to the peer-review process and offer a discussion on future implications of current trends. 
\end{abstract}%%%%%%%%%

\maketitle

\footnotetext{Correspondence:\ 
{robert.west@epfl.ch},
{giuseppe.russo@epfl.ch}}
\footnotetext{Equal contributions, random order.}
\renewcommand*{\thefootnote}{\arabic{footnote}}
\setcounter{footnote}{0}

% Paragraph #1: Importance and challenges of peer reviewing;
\noindent Peer review is central to the modern scientific process and the current epistemic and social status of science~\cite{wilholt2013epistemic,smith2006peer,chubin1990peerless}.
The system is used by journals and conferences to ensure the validity and significance of research findings~\cite{vesper2018peer,ware2015stm} and by funding institutions to allocate grants~\cite{wessely1998peer, PeerReviewGrants, MeritReviewNSF}.
Society treats peer-reviewed research differently from non-peer-reviewed research: it is prioritized by policy advisory groups like the Intergovernmental Panel on Climate Change~\cite{beck2018ipcc,alberts2008reviewing}, holds a special status in the courtroom~\cite{albright2023scientist}, and is often a hard requirement for researchers to progress in their academic career~\cite{sarabipour2019value}.
At the same time, the peer-review system is under mounting pressure~\cite{alberts2008reviewing,tennant2018state}:
the number of researchers~\cite{GlobalStatePeer} and the number of papers published per researcher~\cite{bornmann2015growth} have been growing rapidly, causing the volume of papers that require reviews to outpace the number of qualified reviewers~\cite{ScholarlyPeerReview,arns2014open}. 
This can lead to so-called ``reviewer fatigue''~\cite{breuning2015reviewer} and make recruiting qualified reviewers challenging for some journals~\cite{fox2017recruitment}.

% Paragraph #2: Potential issue \LLM{} reviews
Adding to these problems, the recent emergence and popularization of large language models (LLMs) have raised further concerns within the academic community. 
One key concern is that overburdened scientists~\cite{barnett2019working,geng2022scientists} may resort to using increasingly capable LLMs for peer review~\cite{hosseini2023fighting}. 
While reviews written with LLMs generally resemble ``real'' reviews, reduced reviewer input could lead to the scientific merit of submissions being incorrectly judged~\cite{donker2023dangers}. 
Scientists' reliance on LLMs to write peer reviews (hereinafter called \LLM{} reviews) could thus decrease the peer-review system's reliability and harm its social and epistemic functions~\cite{hosseini2023fighting}.
In response, multiple journals and conferences have already felt obliged to regulate or prohibit the use of LLMs in the peer-reviewing process~\cite{sciencePeerReviewScience,boyd2023acl,icmlICML2023,natureArtificialIntelligenceAI}.

\begin{figure*}[t]
    \centering
    \includegraphics[width=.975\textwidth]{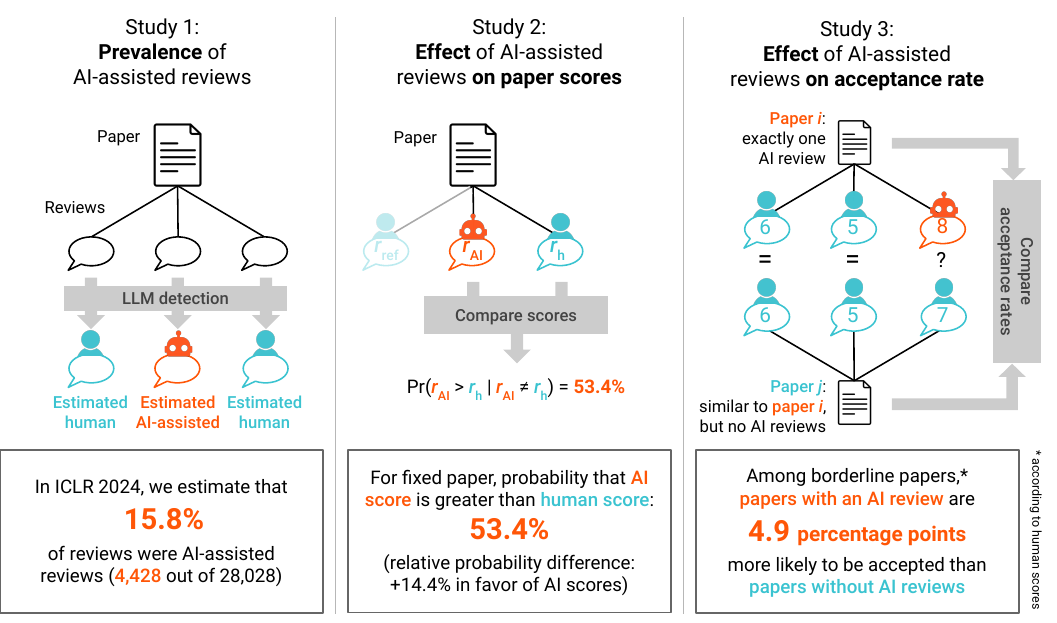}
    \caption{
    \justifying 
    \footnotesize
    \textbf{Overview of our quasi-experimental approach to estimate the prevalence and causal effects of \LLM{} reviews.}
    \textbf{Study 1: Estimating the prevalence of \LLM{} reviews}
    by classifying each review as human or \LLM{} using an out-of-the-box LLM-detection model.
    \textbf{Study 2: Estimating the effect of \LLM{} reviews on paper scores} 
    by comparing the scores of human and \LLM{} reviews assigned to the same paper (thus controlling for properties of the reviewed paper).
    \textbf{Study 3: Estimating the effect of \LLM{} reviews on acceptance rate:} 
    we match papers into pairs $\langle i,j\rangle$ such that 
    (1) $i$ and $j$ are similar in content,
    (2) $i$ and $j$ received the same number $m$ of reviews, 
    (3) $i$ received exactly one \LLM{} review, and $j$ none,
    (4) $i$'s $m-1$ human scores are identical to $m-1$ of $j$'s $m$ human scores. 
    We then estimate the causal effect of \LLM{} reviews on paper acceptance
    as the difference in acceptance rates between $i$ and $j$ in matched pairs. 
    }
    \label{fig:display}
\end{figure*}

% Paragraph #3: Why is it hard?
Despite the increased concerns around \LLM{} reviews, the central question remains unanswered: How do \LLM{} reviews influence peer-review outcomes?
Unfortunately, disentangling the causal effect of \LLM{} reviews is challenging for at least the following reasons:
Firstly, distinguishing between texts generated by LLMs and texts generated by humans is difficult for machine-learning models and humans alike~\cite{sadasivan2023can,jakesch2023human}.
Secondly, even if one can accurately detect their use, it is unclear what role LLMs play in writing \LLM{} reviews: Do they serve as enhanced spell-checkers? Or rather to formulate the core arguments of a review? In the former case, using LLMs may improve the writing quality of reviewers with English as a second language~\cite{amano2023manifold,lin2024techniques}, while in the latter, it may threaten the essence of the peer-review process itself~\cite{donker2023dangers,hosseini2023fighting}.
Finally, even in the pessimistic case where LLMs are used to formulate core arguments, it is unclear whether their impact on paper acceptance decisions is substantial. 
Past work suggests that random chance plays a substantial role in the acceptance of papers into conferences and journals~\cite{esarey2017does,neuripsexp}, which could render the noise added by \LLM{} reviews inconsequential.

% Paragraph #4: ICLR decision process
We address these challenges in the context of a high-profile machine-learning conference, the International Conference on Learning Representations (ICLR).
This conference is unique in adopting an open peer-review model, in which all reviews are visible and easily retrievable, regardless of whether papers have been accepted or not.
The ICLR reviewing process happens in roughly five steps:
(1) reviewers ``bid'' on papers they would like to review based on their expertise and interests, after which an assignment algorithm is run, taking into account reviewers' bids, expertise, potential conflicts of interests, and ``reviewer diversity'' per paper;
(2) papers typically receive three or more reviews that rate the contribution on a scale from 1 to 10 (where 5 and 6 represent borderline scores around the acceptance threshold) and provide a confidence level from 1 to 5; 
(3) authors engage with the reviewers in an asynchronous discussion period; 
(4) reviews are collated and weighted by so-called ``area chairs'' into a meta-review recommending acceptance or rejection; 
(5) ``senior area chairs'' and ``program chairs'' help calibrate the area chairs' recommendations and collectively determine the final decision.

% Paragraph #5: Data and Methods
To estimate the causal impact of \LLM{} peer reviews on submission scores and acceptance rates, we consider all ICLR submissions (\submissionstf) and reviews (\reviewstf) of 2024, extracted through the application programming interface (API) of OpenReview, the platform where ICLR's reviewing process is hosted.
We conduct three studies (overview in Figure~\ref{fig:display}).
In Study~1, we use the commercially available GPTZero LLM detector \cite{tian2023gptzero} to identify reviews that were likely written with the assistance of an LLM (see \MatMeth; our analysis indicates that the model has a low false-positive rate for the data at hand) and quantify the prevalence of \LLM{} reviews. 
We then conduct two quasi-experimental studies to identify the causal impact of receiving an \LLM{} review. 
In Study~2, we estimate the effect of \LLM{} reviews on scores, contrasting \LLM{} reviews with human reviews assigned to the same paper.
In Study~3, we estimate the effect of \LLM{} reviews on acceptance rates, by comparing outcomes within pairs of papers similar in topic, reviews, and scores, but where exactly one of the papers received an \LLM{} review.

We make three main findings.
In Study~1, we find strong evidence that \LLM{} reviews were highly prevalent at ICLR 2024, with at least \estimatedReviewsLLM{} of reviews written with LLM assistance according to the GPTZero LLM detector.
In Study~2, we find that \LLM{} reviews typically increased average submission scores: considering pairs of reviews with different scores assigned to the same paper, \LLM{} scores were higher than human scores in \estimatedEffectScoreIntepretable of pairs (\estimatedEffectScorePVal; relative difference in probability of scoring higher:\ $+$\estimatedEffectScore in favor of \LLM{} reviews). 
In Study~3, we find that \LLM{} reviews boost papers' acceptance rate, especially for submissions with borderline scores:
receiving an \LLM{} review increased the acceptance rate by \estimatedEffectAcceptance (\estimatedEffectAcceptancePval) on average,
and by as much as \estimatedEffectAcceptanceB (\estimatedEffectAcceptanceBPVal) for borderline submissions, corresponding to a \estimatedEffectAcceptanceBOdds relative increase in odds of acceptance (\estimatedEffectAcceptanceBOddsPVal).
In summary, our findings suggest that \LLM{} reviews were widespread at ICLR 2024 and impacted scores and acceptance rates, corroborating concerns that AI use can reduce the utility of and trust in peer-reviewing.
We open-source our code and annotated data allowing other scholars to conveniently replicate and extend our findings:  \url{https://github.com/epfl-dlab/AIReviewLottery}.

\section*{Results}\label{sec:results}

% \vspace{1.25mm}
\xhdr{Study 1: Prevalence of \LLM{} reviews}
We classify each review as \LLM{} or human using GPTZero, a commercial LLM detector~\cite{tian2023gptzero}. 
The final classification was done on 26 April 2024 to ensure that a single API checkpoint (dated 4 April 2024) can be used to reproduce our results.
We label reviews as \LLM{} if the overall probability of the review being human-generated is below 0.5 (we found our analysis robust to this threshold, see \SuppI{}~\ref{si:threshold}).
We repeat this analysis for each year between 2018 and 2024. Since LLMs only became widely available after ChatGPT debuted in November 2022 \citep{chatgptrelease22} (i.e., after the reviewing cycle for ICLR 2023), we use reviews from 2018 to 2023 to estimate GPTZero's false-positive rate (FPR) and correct the 2024 estimate by removing the average FPR of previous years from the 2024 prevalence estimate (see \MatMeth). We do not estimate or correct for GPTZero's false-negative rate, meaning that the results provided here are a lower bound of actual LLM prevalence. 

\begin{figure}[t]
    \centering
    \includegraphics[scale=.75]{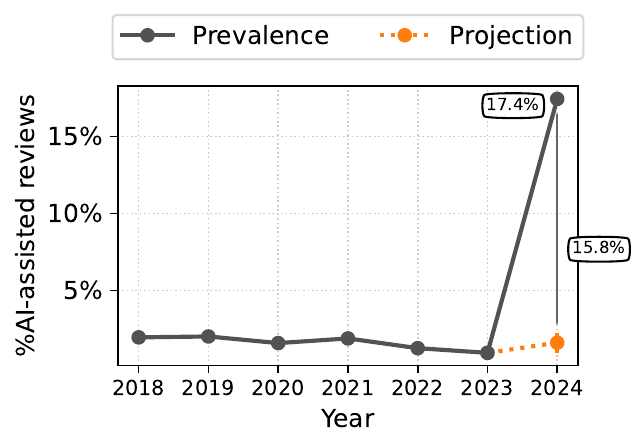}
    \caption{\footnotesize \justifying \textbf{Estimated prevalence of \LLM{} ICLR reviews 2018--2024 (Study~1).} Using the LLM detector's predictions in pre-ChatGPT years (2018--2023) to calculate its false-positive rate, we estimate that \estimatedReviewsLLM{} of reviews in 2024 were \LLM{} (prevalence minus projection in the plot). We estimated 95\% confidence intervals using bootstrap resampling for the prevalence (gray line), but they are too small to be visible. For the projection (orange line; the average prevalence between 2018 and 2022), we plot an error bar corresponding to the prevalence ranges observed in previous years.}
    \label{fig:rq1}
    \vspace{-1mm}

\end{figure}

With this method, we estimate that \estimatedReviewsLLM{} of ICLR reviews in 2024 were crafted with the assistance of an LLM, or 4,428 of the 28,028 reviews submitted that year; 49.4\% of all submissions received at least one review classified as \LLM{} by GPTZero.
Figure~\ref{fig:rq1} illustrates the fraction of reviews classified as \LLM{} across the years.
These results are consistent with those of concurrent analyses \cite{liang2024mapping, liang2024monitoring} that used a different methodology.

% \vspace{1mm}
\xhdr{Study 2: Effect of \LLM{} reviews on paper scores} 
Having determined that \LLM{} reviews were common in ICLR 2024, we next estimate their causal effect on paper scores. As illustrated in Figure~\ref{fig:display} (Study~2), we focus on submissions with at least one \LLM{} review (according to GPTZero, per Study~1) and at least two human reviews (\submissionsAItfone submissions).
For each such submission with at least three reviews, we let the score of one of the human reviews be the reference score ($r_{\text{ref}}$) and estimate the difference between the score of an \LLM{} review ($r_{\text{ai}}$) and another human review ($r_{\text{h}}$)
Note that each paper has multiple possible combinations of human, \LLM{}, and reference reviews. We consider all combinations ($n=9{,}666$) and ensure the validity of our results by weighting analyses such that each paper contributes equally to the results and using robust standard errors clustered at the paper level.
Overall, our setup compares pairs of reviews assigned to the same paper, which controls for paper-level confounders, e.g., that specific topics might attract higher- or lower-quality reviews.

\begin{figure}[t]
    \centering
\includegraphics[scale=.75]{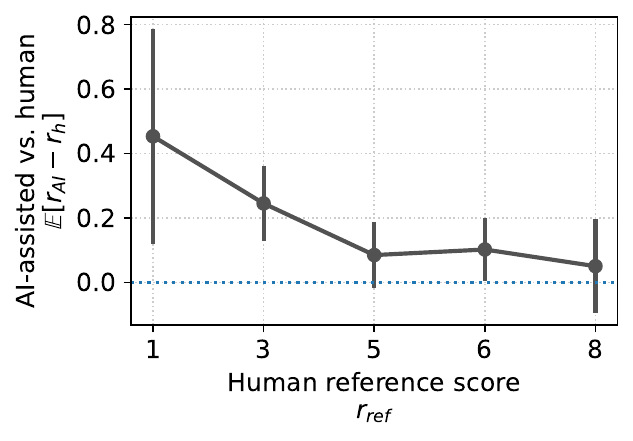}
    \caption{\justifying \footnotesize \textbf{Mean submission-level differences between \LLM{} and human reviews as a function of human reference scores (Study~2).} 
    We consider submissions with at least three reviews, where at least one is \LLM{} and at least two are human. 
    Then, we select a human review as the reference review (with score $r_\text{ref}$) and estimate the average difference between \LLM{} and human reviews ($r_\text{AI} - r_\text{h}$). 
    In the plot, we show the average difference ($y$-axis) for each possible score of the reference review ($x$-axis).
    \LLM{} reviews consistently give higher scores than human reviews.}
    \label{fig:rq2}
    \vspace{-1mm}
\end{figure}

We find that, on average, \LLM{} reviews were 0.14 points (95\% CI [0.08, 0.19]) higher than human reviews.
In Figure~\ref{fig:rq2}, we plot the average difference between human and \LLM{} scores of the same submission ($y$-axis) as a function of human reference scores ($x$-axis). 
We note that \LLM{} reviews consistently assign higher scores than human reviews. For instance, when the score of the human reference review equals~1 (lowest possible score), \LLM{} reviews tend to score submissions 0.45 (95\% CI [0.13, 0.78]) points higher than human reviews do.
Given that scores are ordinal rather than scalar, we complement the previous result with an ordinal regression analysis~\cite{liddell2018analyzing}. 
Using a proportional odds model~\cite{mccullagh1980regression}, we regress the score of reviews as a function of an indicator variable coding whether the review was \LLM{}.
We estimate that if we select two reviews such that they have different scores and such that exactly one is \LLM{}, the probability that the review with the higher score is the \LLM{} one equals \estimatedEffectScoreIntepretable (relative probability difference:\ $+$\estimatedEffectScore; \estimatedEffectScorePVal).

\xhdr{Study 3: Effect of \LLM{} reviews on acceptance rate}
Although we have shown that \LLM{} reviews boost submissions' average scores, this does not automatically translate into better chances of acceptance. 
For instance, it could be that predominantly submissions with very low or high average scores receive \LLM{} reviews, which would likely render the boosted average scores inconsequential for acceptance. 
Crucially, simply comparing the average acceptance rate of submissions that received \LLM{} reviews with those that did not is insufficient to estimate the causal effect of AI assistance. For example, submission- and review-related features might confound treatment (receiving an \LLM{} review) and outcome (being accepted).
To control for possible confounders, we thus isolate the effect of \LLM{} reviews on acceptance in a matched study, where we compare submissions that are similar or identical in a wide range of aspects, but one received an \LLM{} review and the other did not.

Matching is done in two steps.
Firstly, we select submissions that received exactly one \LLM{} review.
For each such submission $i$ we then curate a set of possible matches consisting of submissions that received the same number of reviews as $i$,
all of which are classified as human,
and all but one of which have scores identical to the scores of $i$'s human reviews.
For example, if a submission received two human reviews with scores of 6 and 5 and an \LLM{} review with a score of 8, a possible candidate might have received three human reviews with scores of 6, 5, and 7 (see Figure~\ref{fig:display}, Study 3).
Secondly, we rank these possible matches by measuring their semantic similarity with~$i$ (embedding abstracts and review content with Sentence-BERT~\cite{reimers2019sentence}; see \MatMeth) and choose the best-matching candidate as $i$'s match $j$.

Considering all submissions in this matched sample (\matchedSampletf), we estimate the effect of receiving an \LLM{} review on the acceptance of a submission $k$ by fitting a logistic regression
\begin{equation}\label{eq:acceptance}
    \text{logit}(y_k) = \alpha + \beta \cdot L_k + \mathbf{\gamma} \cdot \mathbf{X}_k,     
\end{equation}
where $L_k \in \{0,1\}$ indicates whether $k$ had an \LLM{} review and $\mathbf{X}_k$ is a vector with the same control variables used for matching. We also estimate the equivalent linear regression
\begin{equation}
    y_k = \alpha + \beta \cdot L_k + \mathbf{\gamma} \cdot \mathbf{X}_k.
\end{equation}
In both regressions, $\beta$ captures the difference in acceptance (in log odds for the logistic regression and percentage points for the linear regression) between submissions receiving vs.\ not receiving an \LLM{} review, ceteris paribus. 

The fitted coefficients reveal that submissions that received \LLM{} reviews had \estimatedEffectAcceptanceOdds higher odds of being accepted compared to those that did not (\estimatedEffectAcceptanceOddsPVal), or alternatively, had \estimatedEffectAcceptance higher chances of being accepted (\estimatedEffectAcceptancePval).
This average effect, however, downplays the potential impact of \LLM{} reviews, as submissions whose other reviews have very low or very high scores are less likely to have their acceptance decision flipped by the \LLM{} review.

\begin{figure}[t]
    \centering
    \includegraphics[scale=0.75]{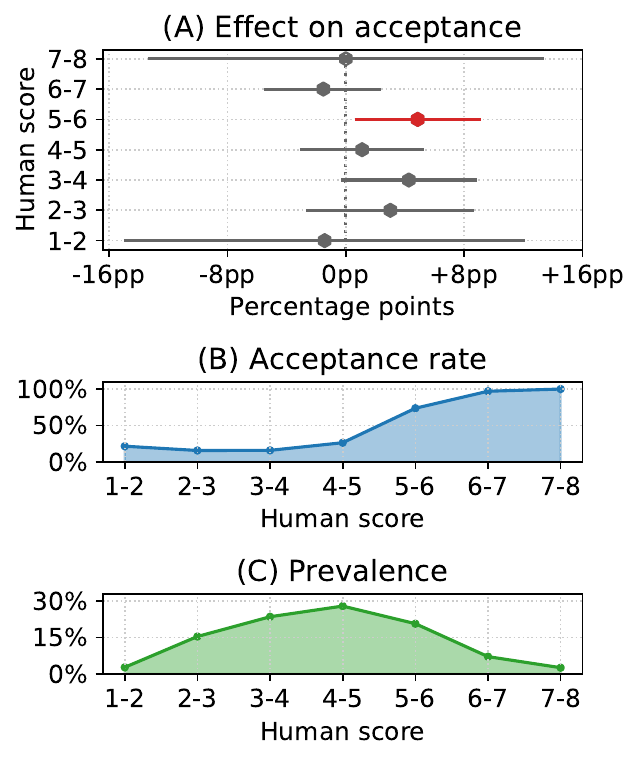}
    \caption{
    \footnotesize \justifying 
    \textbf{Effect of receiving an \LLM{} review on submission acceptance (Study~3).} 
    (A) We stratify the effect of \LLM{} reviews on submission acceptance by matched submissions' average score across the human reviews they received ($y$-axis). We find a particularly pronounced effect for ``borderline'' submissions (average score between 5 and 6), with an increased acceptance rate of \estimatedEffectAcceptanceB percentage points (\estimatedEffectAcceptanceBPVal). Overall, we find that submissions that received an \LLM{} review are \estimatedEffectAcceptance percentage points more likely to be accepted (\estimatedEffectAcceptancePval).
    (B)~Acceptance rate and
    (C)~prevalence of submissions for submissions receiving only human reviews across human-score bins. E.g., \prevalenceB of submissions were in the $[5, 6)$ bin, and submissions receiving only human reviews in this bin were accepted \acceptanceRateB of the times.}
    \label{fig:rq3}
    \vspace{-1mm}
\end{figure}

We thus study the heterogeneity of the effect by stratifying the matched sample (see Figure~\ref{fig:rq3}(A)).
Each matched pair of submissions with $m$ reviews shares $m-1$ human review scores. 
We take the average value among these $m-1$ review scores and place each pair of matched submissions in one of seven bins, $[1,2), [2,3), \dots, [7,8)$ (see $y$-axis of Figure~\ref{fig:rq3}(A)). 
For example, the matched submissions in Figure~\ref{fig:display} (Study 3) would be placed in the $[5,6)$ bin since the average human score among the review scores they share is $5.5$ (see \MatMeth for details on the matching).
We then repeat the regression for the matched submissions in each bin, finding that the effect is especially pronounced for borderline submissions, i.e., those in the $[5,6)$ bin. Note that per the reviewer instructions, scores 5 and 6 correspond to slightly below and slightly above acceptance. 
More precisely, among these borderline submissions, the acceptance rate of submissions with at least one \LLM{} review is \estimatedEffectAcceptanceB (\estimatedEffectAcceptanceBPVal; linear regression) higher than that of submissions with only human reviews, corresponding to a \estimatedEffectAcceptanceBOdds (\estimatedEffectAcceptanceBOddsPVal; logistic regression) relative increase in the odds of being accepted. 
This is substantial, as the $[5,6)$ bin contains \prevalenceB of all submissions in the matched sample. 
Further, with a \acceptanceRateB acceptance rate in the $[5,6)$ bin (for submissions without \LLM{} reviews in the matched sample), the \estimatedEffectAcceptanceB of absolute increase corresponds to a 6.5\% relative increase in the chance of being accepted. (For a sensitivity analysis, see \SuppI{}~\ref{si:sensitivity}.)

% Our results rely on the assumption that the \LLM{} submissions are comparable to the human submissions in the matched sample. 
% This might be violated if important covariates were not measured, such that submissions that appear comparable may, in fact, not be comparable \cite{rosenbaum2005sensitivity}. 
% To address this concern, we conducted a sensitivity analysis on the effect of \LLM{} on borderline papers that analyzed how strong the effect of an unmeasured covariate would need to be to alter the conclusion of this study, finding that the results are robust against possible unmeasured covariates (see \SuppI{}~\ref{si:sensitivity}). 

\section*{Discussion} \label{sec:discussion}
\noindent 
We studied whether \LLM{} reviews affected the peer-review process of the ICLR 2024 machine learning conference by (1) estimating their prevalence; 
(2)~comparing the scores of \LLM{} vs.\ human reviews for the same submission; 
and (3) comparing the acceptance rate of submissions that received \LLM{} reviews to that of similar submissions that did not. 
Our results suggest that 
(1) LLMs were widely used in the peer-reviewing process of ICLR 2024; 
(2) receiving an \LLM{} review inflated submission scores; and
(3) \LLM{} reviews boosted acceptance rates, especially for borderline submissions.

These findings have important ramifications.
They raise the concern that, in an already overloaded peer-review system, \LLM{} reviews can reduce trust in the process---and science as a whole---by introducing a new point of failure.
This may weaken the epistemic status of a system already deemed ``unscientific'' by some~\cite{rennie2016let}. 
Scientific works that express views relating to societal norms and values may be at even greater risk due to the known biases present in LLMs~\cite{hartmann2023political}, which may reward research that aligns with the implicit values of the LLMs used by reviewers.
As the landscape of LLMs and LLM usage changes, our findings highlight the urgent need to establish baselines and ongoing measurements, accompanying the co-evolution of LLMs and peer-reviewing.

Earlier, we identified three key trends likely pushing reviewers to resort to AI assistance: the increasing submission volume, the dwindling reviewers-to-submissions ratio, and the improving quality of LLM tools.
The pressure induced by these trends will be amplified for more senior peer-review roles, such as (senior) area chairs, who might thus be similarly tempted to lean on AI assistance for writing meta-reviews and making editorial decisions.
Yet, unlike the first peer-review layer, which consists of multiple reviews, meta-reviews present a single point of failure. 
Furthermore, they are often shorter and contain less forced structure, simplifying the usage and complicating the detection of AI assistance.
Given the revealed prevalence of \LLM{} reviews, a more subtle consequence of \LLM{} editorial decisions could emerge: 
there has been increasing evidence that LLMs exhibit a preference toward their own outputs~\cite{bai2024benchmarking, zheng2024judging, llm-self-pref}, which could influence decisions to favor \LLM{} reviews disproportionally.
This preference towards LLM-generated work could even lead authors to ``game'' the system by writing text that aligns with popular model preferences, e.g., tailoring the submission's content to receive better (automated) reviews or injecting special (hidden) instructions into manuscripts.

Despite this study's contribution to quantifying the causal effects of \LLM{} reviews, it is limited in several ways.
Firstly, the methodology proposed here requires both accepted and rejected submissions, alongside the scores of the individual reviews. 
ICLR is one of the few scientific venues that provide conditions for external researchers to carry out our methodology.
Nevertheless, we note that many of the top AI conferences' reviewer pools are very similar to ICLR's. 
For example, the reviewer pools of ICLR 2024 and the 2023 Conference on Neural Information Processing Systems (NeurIPS) had 53.4\% overlap (NeurIPS is another prestigious AI conference; see \SuppI{}~\ref{si:overlap} for overlap comparisons to other conferences).
We thus conjecture that these strongly related conferences might be subject to similar dynamics as ICLR 2024.

Secondly, we estimate the causal effect of \LLM{} reviews on submission-related outcomes, which differs from estimating the effect of \LLM{} reviews on the quality of the reviews. 
It could be that the quality of reviews remains essentially unchanged, and what changes is their delivery: shorter reviews that previously did not meaningfully engage with a submission may have been substituted by more verbose, eloquently written LLM reviews.
Assessing the quality of \LLM{} reviews constitutes an important avenue for future research.

Thirdly, although we have established that \LLM{} reviews were distinct from human reviews assigned to the same or similar papers, our results provide limited insights into how LLMs and other AI tools are used in peer-reviewing. It could be that two modes of usage are prevalent, e.g., improving the text of a previously self-written review vs.\ feeding the paper that is to be reviewed to an LLM and copying the LLM-written review verbatim, and that only the latter impacts peer-reviewing outcomes.
Understanding nuances around these usage modes is of key importance for future decision-making. Yet, it may require a different register of research methods from those deployed here, based on directly engaging with reviewers via interviews, focus groups, and surveys.

It is important to emphasize that using LLMs in reviewing may not be categorically wrong. 
There are various areas where LLMs may improve the current peer-review process.
For example, LLMs could offer reviewers feedback to improve writing clarity, 
detect flawed critiques to reduce misunderstanding, 
or help contextualize the importance of a submission's findings. 
They might even provide new ways to tackle problems poorly addressed by the current peer-review process, e.g., by conducting automated tests to alleviate the ``reproducibility crisis''~\cite{kapoor2023leakage}.
Nonetheless, moving forward, it seems imperative to rethink reporting requirements at every level to ensure the integrity, validity, and transparency of peer review. 
Defining requirements and formulating unified guidelines for the integration of LLMs in peer review
will require community participation. 
Similar to a workshop on peer review held at the 2012 International Conference on Machine Learning (ICML)~\cite{soergel2013open} or the quadrennial International Congress on Peer Review and Scientific Publication~\cite{prc}, it may again be time to organize a workshop discussing the future of peer review.

\section*{Materials and Methods}
\label{sec:material-n-methods}

\xhdr{Dataset} \label{sec:material-n-methods:dataset}
We analyze submission and review data from ICLR, a leading venue in machine learning that publicly releases all peer reviews and decisions after concluding the peer-review process. 
Data extraction was done with the official OpenReview API and consists of all conference submissions and reviews from 2018 to 2024. The final dataset comprises $23{,}959$ main conference submissions and $86{,}690$ reviews, each with overall ratings, textual explanations, and confidence scores.
In addition to a written review, each review at ICLR contains an overall ordinal rating $r \in \{1, 3, 5, 6, 8, 10\}$. Scores 1 and 3 indicate low-quality submissions that should typically be rejected; scores 5 and 6 indicate borderline submissions; and scores 8 and 10 indicate high-quality submissions that should typically be accepted. 
Each score is accompanied by a short textual description, e.g., the description for score 5 reads ``Marginally below the acceptance threshold.'' We depict descriptions for all scores in the \SuppI{}~\ref{si:data}.

\xhdr{Prevalence of \LLM{} reviews} 
We use GPTZero, a commercially available LLM detector.
For each review, GPTZero calculates the probability of it being entirely human-generated, entirely AI-generated, or ``mixed.''
We label reviews as \LLM{} if the probability of the review being human-generated is below 0.5 (for analyses with varying threshold values, see \SuppI{} \ref{si:threshold} to ensure robustness). 
Accordingly, reviews with a human-generated score below 0.5 indicate a cumulative probability of being entirely AI-written or ``mixed'' greater than 0.5. 
Using reviews written before ChatGPT's popularization, we estimate that the model's false positive ratio (FPR) is \fpratev for the data studied. 
Under the assumption that the FPR remains the same for 2024 reviews, we estimate a lower bound of the overall prevalence of AI assistance in 2024 by subtracting the FPR from the fraction of reviews classified by GPTZero to be \LLM{}. 
This is similar to well-established methods to prevent misclassification bias~\cite{meyer2017misclassification}. It is important to note that GPTZero was not trained on pre-ChatGPT submission reviews as human text (Alex Cui, CTO of GPTZero, personal communication, 22 April 2024), which could bias the FPR estimate and thus our results. 

\xhdr{Proportional odds model} \label{sec:material-n-methods:ppo}
Since ratings are ordinal, using a linear regression model to estimate score differences can lead to systematic errors as the response categories of an ordinal variable may not be equidistant~\cite{liddell2018analyzing, burkner2019ordinal}. 
A solution to this issue is using ``cumulative'' ordinal models that assume that the observed ordinal variable comes from categorizing a latent, non-observable, continuous variable~\cite{burkner2019ordinal}. 
Here, we use one such model, a ``proportional odds model'' of the form
 \begin{equation}
     \label{eq:rq2}
     \log \frac{\Pr(r_k \leq a)}{\Pr(r_k>a)} = \alpha_a - \gamma \cdot L_k,
 \end{equation}
where $a \in \{1, 3, 5, 6, 8\}$ represents the possible values the review $k$ might take; $\beta_a$ and $\beta_{Aa}$ are level-specific coefficients; $R$ is the set containing all possible review scores; $L_k$ is an indicator variable that equals 1 when the review is \LLM{};
Under this specification, $e^\gamma$ corresponds to the odds ratio
\begin{equation}
     \label{eq:rq2b}
    e^\gamma = \frac{\frac{\Pr(r > a | L)}{\Pr(r \leq a | L)} 
    }{ \; \frac{\Pr(r > a | \neg L)}{\Pr(r \leq a |  \neg L)}\;}, \;\;\; \text{for all } a \in \{1, 3, 5, 6, 8\}.
\end{equation}
While the regression estimates the odds ratios, we also present results in an equivalent but easier-to-interpret fashion. Suppose we pick two reviews with different scores assigned to the same submission, exactly one of which is \LLM{}.
The odds ratio $e^\gamma$ estimated by the model corresponds to the odds $\kappa$ that the \LLM{} review scores higher in the paired scenario delineated above.
Note that $\kappa = x / (1-x)$, where $x$ is the probability of the \LLM{} review having a higher score. This $x$ is the number we report in the paper, e.g., \estimatedEffectScoreIntepretable in Figure~\ref{fig:display}.
See \SuppI{} \ref{si:score} for the full regression table.

\xhdr{Matching}
We estimate the effect of \LLM{} reviews on submissions' acceptance employing a two-step matching procedure.
For each submission that received exactly one \LLM{} review, we first construct a set of submissions that have the same number of reviews, no \LLM{} reviews, and matching human-review scores for all but the \LLM{} review.
Then, we use content-based matching to determine the best match among the candidate set. For each submission and set of submission candidates, this involves: (1) computing the embeddings for the abstract and the content of reviews associated with the submission using Sentence-BERT~\cite{reimers2019sentence}, 
(2) concatenating these embeddings into a single vector, 
and (3) measuring the cosine similarity between the vector of the submission that received the \LLM{} review and each candidate submission.

We select the candidate with the highest cosine similarity.
We only keep matches with a cosine similarity above a threshold of 0.1 (the effects of changing this threshold are discussed in the \SuppI{} \ref{si:accept}). This process matches 98.5\% of potential pairs, or 2,580 out of 2,619 submissions.
Additionally, we conducted several checks to ensure the quality of our matches. 
We first compared the content similarity between the matched sample and a sample of ``randomly'' matched submissions. This unmatched sample consists of the matched sample's submissions that received an \LLM{} review and a randomly sampled submission that did not receive an \LLM{} review. 
We compare the overlap of the keywords (as submitted by the authors) used in the matched sample with the overlap of the unmatched sample. 
We observe a significantly higher overlap for the matched sample. 
We further compared the similarity between the Sentence-BERT embeddings of the abstracts within the matched sample. These similarity scores were statistically compared to those of the unmatched sample to confirm a higher and statistically significant difference. We provide more details on these analyses in \SuppI{}~\ref{si:accept:balance-matching} and a sample of matched abstracts in \SuppI{}~\ref{sec:si:examples}. 

\xhdr{GPTZero robustness checks}
Given that our analysis is based on GPTZero predictions of LLM-generated text,  we further assess the reliability of this classifier.
We construct a vocabulary consisting of all words used in ICLR 2024 reviews.
Then, for each word in this vocabulary, we compute the ratio between the number of LLM-generated reviews containing the word and the number of human reviews containing the word.
Among the words with the highest ratio, we find words identified by other work as indicative of LLM use ~\cite{liang2024mapping, liang2024monitoring}, e.g., ``delve'', ``bolster'', and ``illustrates''; see \SuppI{}~\ref{si:prevalence} for more details.
Beyond this word-level ratio check, we vary the adopted threshold of 0.5 used to label a review as \LLM{}. 
Our analysis detailed in \SuppI{}~\ref{si:threshold} shows that our results remain robust across different values of this threshold. 

\clearpage
\onecolumngrid

% \subsubsection*{Acknowledgements}
% \begin{center}
%     \textit{Acknowledgments}
% \end{center}
\xhdr{Acknowledgments}
The authors would like to thank Alex Cui and the GPTZero team for their generous support of our work. 
Robert West's lab is partly supported by grants from the Swiss National Science Foundation (200021\_185043, TMSGI2\_211379), Swiss Data Science Center (P22\_08), H2020 (952215), Google, and Microsoft.

\bibliographystyle{plain}
\bibliography{refs}

\clearpage

\appendix
\renewcommand{\thesection}{\Alph{section}}

\begin{center}
    \Large
    \SuppI{}
\end{center}

\vspace*{\fill}

\section{Dataset}\label{si:data}

\xhdr{Submissions}  \label{si:data:submissions}
When a submission is made to ICLR and it is not withdrawn before the submission deadline, it is hosted on OpenReview. As mentioned in the section \MatMeth{}, we use the OpenReview API to collect all submissions made to ICLR between 2018-2024, including the titles, abstracts, introductions, keywords, and author institutions. This collection resulted in 23,959 main conference submissions from 46,257 authors and 1,263 institutions (see Table~\ref{tab:summary}).

\input{tables/dataset}

\xhdr{Reviews}\label{si:data:revies}
Similarly, we collect all 86,690 reviews spanning from 2018 to 2024 (see table \ref{tab:summary}). These reviews include textual evaluations, confidence scores, and overall score ratings. 
The confidence score was measured using a consistent scale of one to five during the observation period. 
The overall score rating scale consists of a one to ten scale but with only specific scores possible (1, 3, 5, 6, 8, 10). Each of these scores comes with a detailed definition to provide guidance on their meaning:
\begin{itemize}
    \item 1: strong reject.
    \item 3: reject, not good enough.
    \item 5: marginally below the acceptance threshold.
    \item 6: marginally above the acceptance threshold.
    \item 8: accept, good submission
    \item 10: strong accept, should be highlighted at the conference.
\end{itemize}

\vspace*{\fill}

\clearpage

\vspace*{\fill}

\section{On the Robustness of the \LLM{} Labeling Threshold}\label{si:threshold}

\begin{figure}[ht]
    \centering
    \includegraphics[scale=.65]{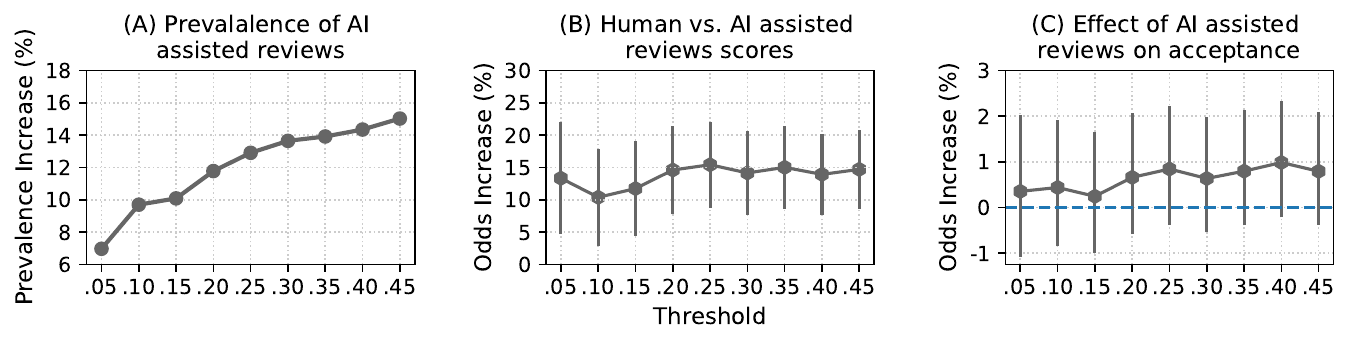}
    \caption{\footnotesize \justifying \textbf{Robustness of the \LLM{} reviews labeling threshold.} The plots show the robustness of the 0.5 threshold used to label reviews as \LLM{} or human. The plots show the prevalence analysis (A), the reviews score difference analysis (B), and the acceptance analysis (C) when varying the threshold.} 
    \label{fig:sensitivity_panel}
\end{figure}

\noindent 
We label reviews as \LLM{} if GPTZero predicts their probability of being human written as less than 0.5. 
Here, we explore the robustness of our findings in the three studies conducted (prevalence, \LLM{} vs. human scores, and acceptance analysis) by varying this threshold. 
Specifically, we select the following threshold values: [0.05, 0.10, 0.15, 0.20, 0.25, 0.30, 0.35, 0.40, 0.45]. 
We choose thresholds \textit{lower} than 0.5 because higher thresholds would yield higher false-positive rates and, thus, higher bias in the estimates (as human reviews would be classified as \LLM{}). 
We depict results in Figure \ref{fig:sensitivity_panel}, confirming that our findings remain valid when changing the labeling threshold. 

\xhdr{Prevalence Analysis} Figure \ref{fig:sensitivity_panel}(A) shows the prevalence of \LLM{} reviews corrected for the false positive rate, as done in the primary analysis. 
Interestingly, even under a stringent threshold---classifying a review as \LLM{} if the probability of being human is less than 0.05---we still find that over 7\% of reviews submitted to ICLR are \LLM{}.

\begin{figure}[ht]
    \centering
    \includegraphics[scale=.55]{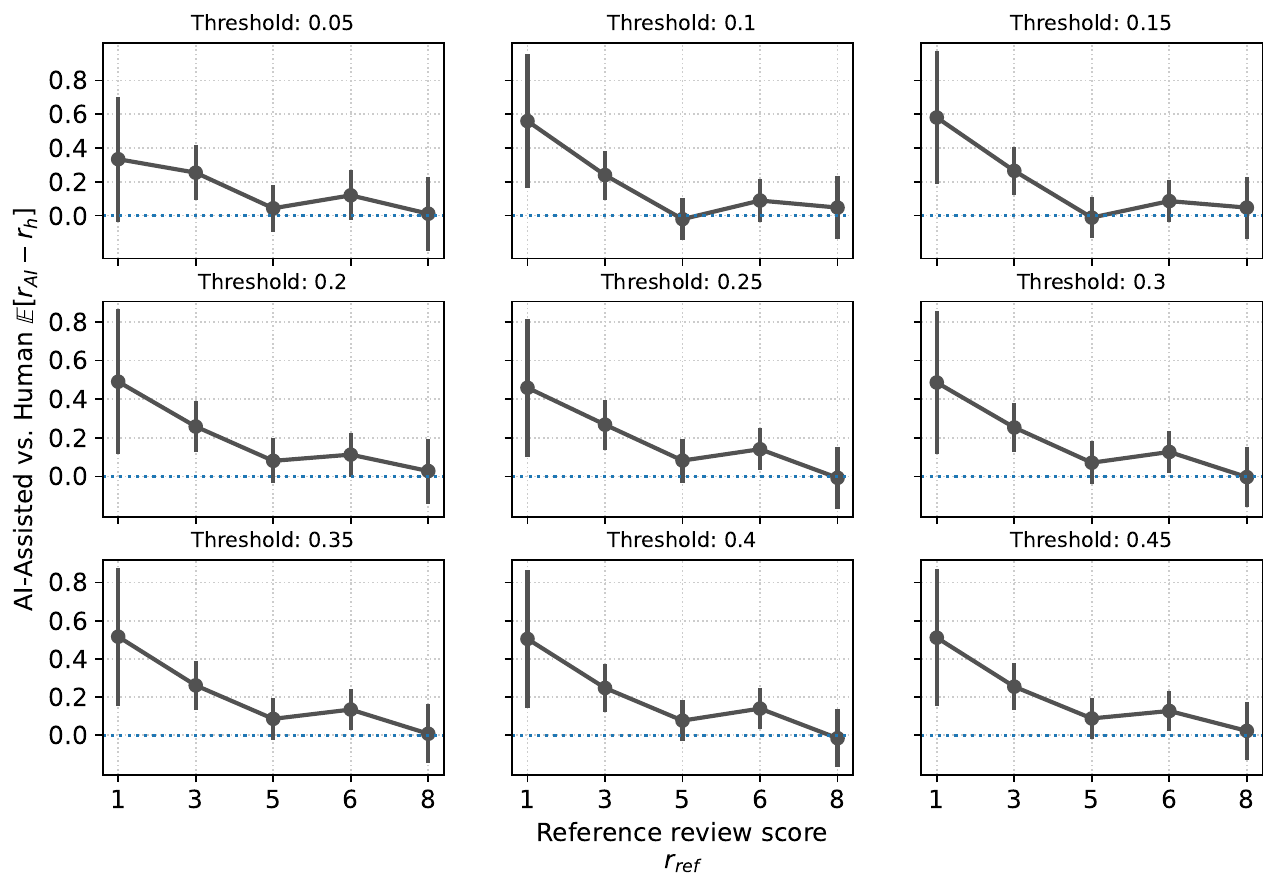}
    \caption{\footnotesize \justifying \textbf{Mean submission-level differences between \LLM{} and human reviews, as a function of human
reference scores for all labeling thresholds considered.} This figure reproduces Figure~\ref{fig:rq2} using different thresholds to label reviews as \LLM{} or human.} 
    \label{fig:score_sensitivity}
\end{figure}

\xhdr{Reviews Scores Difference Analysis} 
Figure \ref{fig:sensitivity_panel}(B) shows the increase in the odds of receiving a higher score from an \LLM{} review over a human review across thresholds, calculated using the proportional odds model (see Equation~\ref{eq:rq2}).
We further reproduce Figure~\ref{fig:rq2} using different thresholds in Figure~\ref{fig:score_sensitivity}, showing that the findings are robust to the picked threshold. 

\begin{figure}[ht]
    \centering
    \includegraphics[scale=.55]{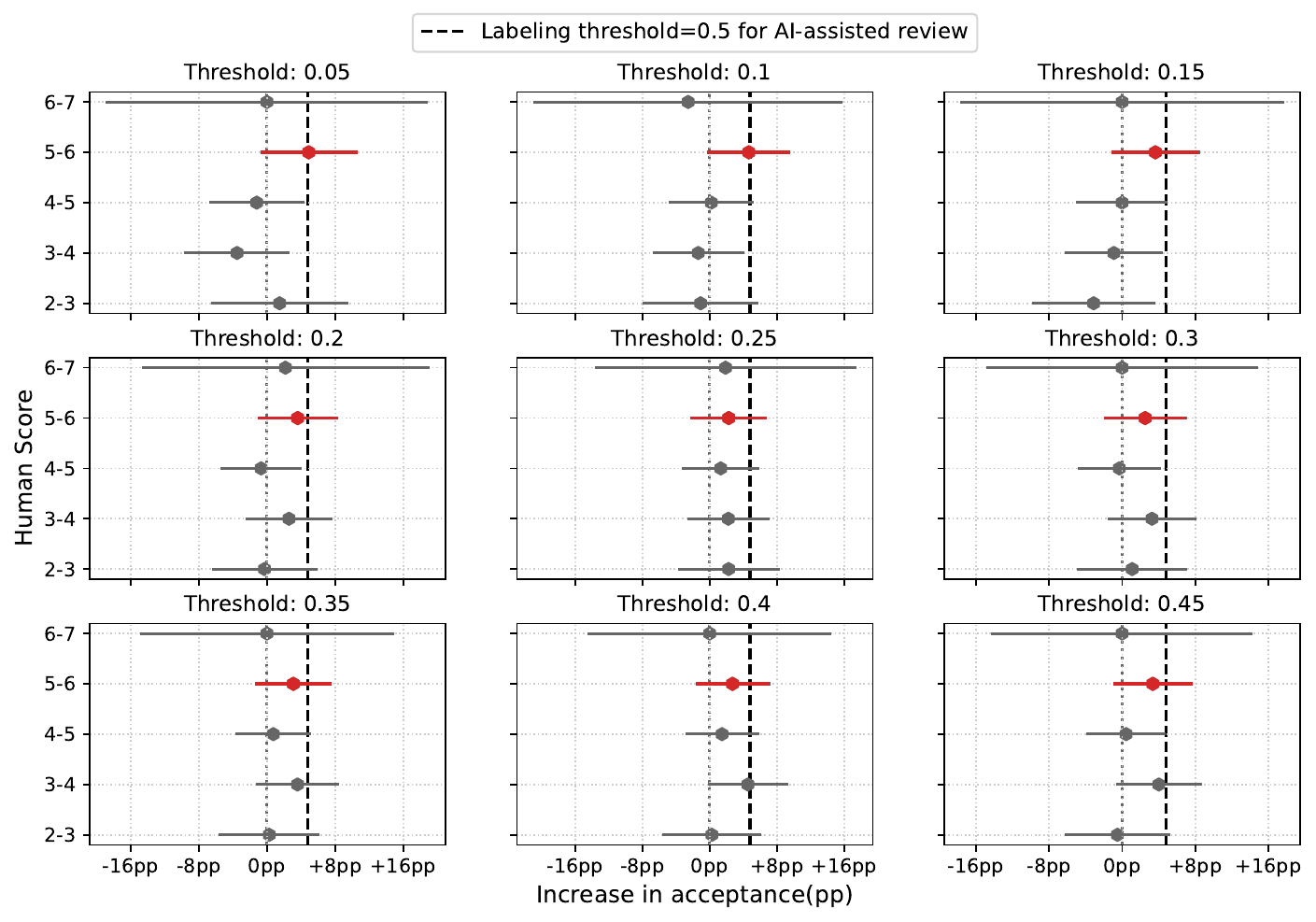}
    \caption{\footnotesize \justifying \textbf{Effects of receiving an \LLM{} review on acceptance.} This figure reproduces Figure~\ref{fig:rq3}(A) using different thresholds to label reviews as \LLM{} or human.} 
    \label{fig:acceptance_sensitivity}
\end{figure}

\xhdr{Acceptance Analysis} Figure \ref{fig:sensitivity_panel}(C) shows the average increase in odds across thresholds, calculated using the logistic regression depicted in Equation~\ref{eq:acceptance}. The effect sizes remain consistent, confirming the stability of our threshold settings. 
We further reproduce Figure~\ref{fig:rq3}(A) using different thresholds in Figure~\ref{fig:acceptance_sensitivity}, showing that the findings are robust to the picked threshold.

\vspace*{\fill}

\clearpage

\vspace*{\fill}

\section{On the Effect of LLMs on Review Scores }\label{si:score}
\input{tables/scores_ordinal_regression}

\vspace*{\fill}

\clearpage

\vspace*{\fill}

\section{On the Effect of LLMs on Submissions Acceptance}\label{si:accept}

\xhdr{Assessing Matching}\label{si:accept:balance-matching}
The matching procedure used to estimate the effect of \LLM{} reviews on acceptance rates consists of two steps: 
(1) an exact match based on the number of reviews and the scores assigned to the submissions, 
and (2) a content-based match computed using the cosine similarity of the embeddings of abstracts and content of their associated reviews. We only include matches in our analysis where the cosine similarity exceeds 0.1 (we refer to this value as a matching threshold).

\begin{figure*}[ht]
    \centering
\includegraphics[scale=0.6]{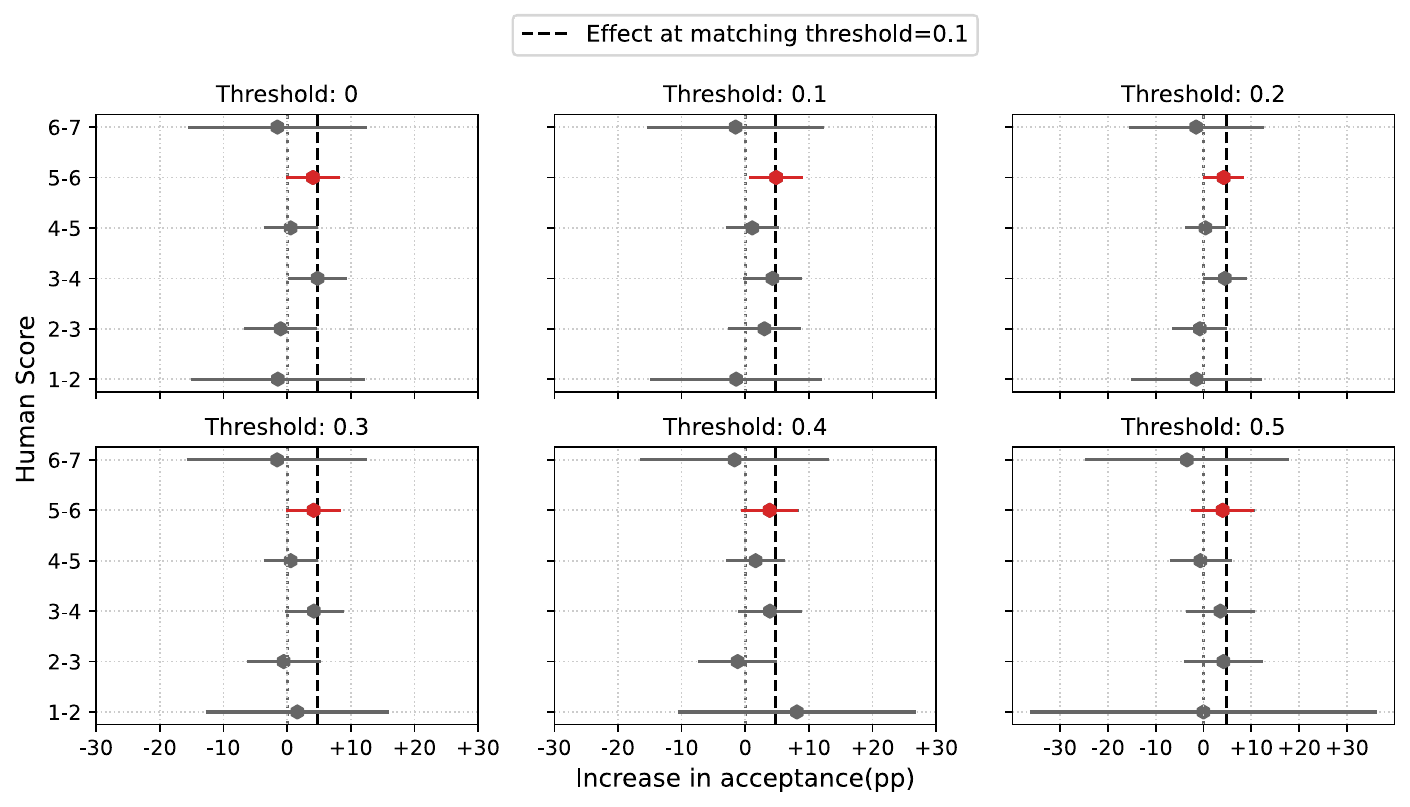}
    \caption{\footnotesize \justifying \textbf{Sensitivity of the matching threshold.} We show how the stratified effects for the acceptance analysis change when varying the matching threshold of 0.1. 
    Effect sizes remain qualitatively similar to the one estimated in our main analysis (black dash line) showing the robustness of our matching.}
    \label{fig:matching_threshold}
\end{figure*}

\begin{figure*}[ht]
    \centering
\includegraphics[scale=0.6]{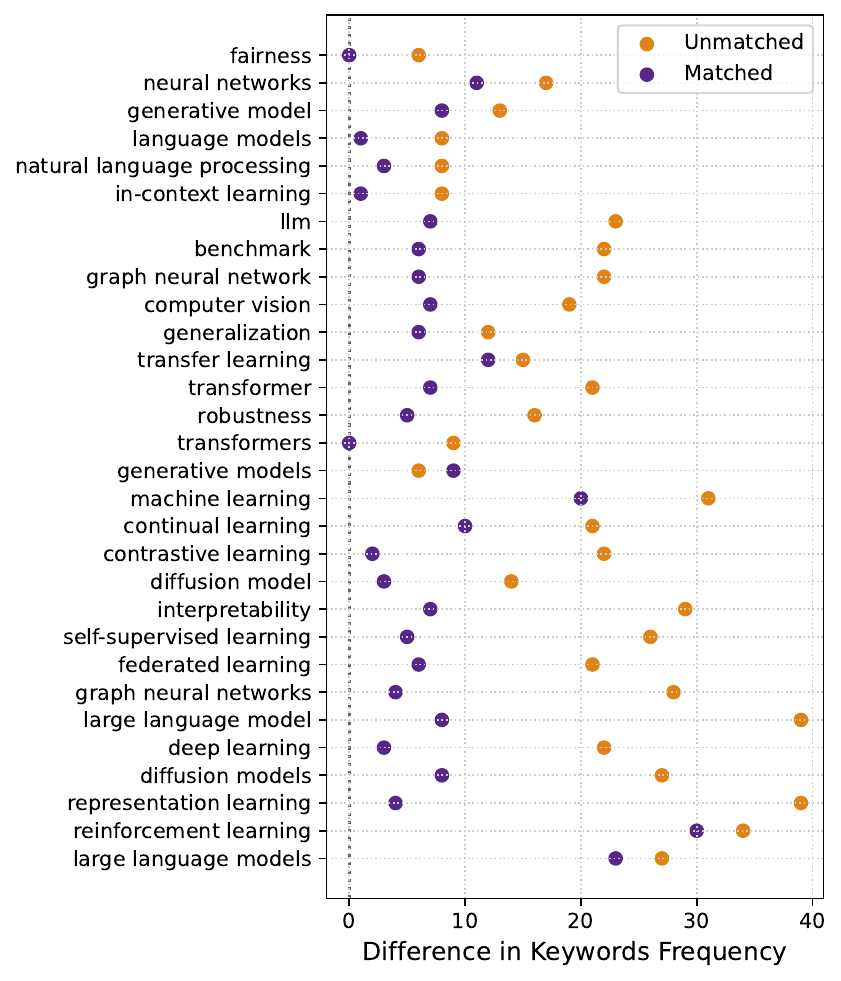}
    \caption{\footnotesize \justifying \textbf{Difference in keywords frequency.} In purple we show the difference in keywords frequency for the matched (purple) and unmatched sample (orange). We consistently observe that the matched submissions consistently have lower topic discrepancies compared to the unmatched sample. }
    \label{fig:keywords_balance}
\end{figure*}

Figure \ref{fig:matching_threshold} shows how the impact of receiving an \LLM{} review on the acceptance rates of ``borderline'' papers remains consistent across various matching thresholds in our stratified analysis. The effect is similar to what we observed in our main analysis and shows no statistical difference, confirming the reliability of our matching approach.
We assess the quality of our matched sample by comparing the content similarity between matched submissions and a sample of ``unmatched'' submissions (for each submission in the matched sample, we randomly sample a submission from ICLR 2024). We used these samples in two robustness checks:

\begin{itemize}
    \item \textit{Analysis of Keywords Similarity} 
    For each sample, we checked for keyword overlap in each pair, hereinafter ``hits''. The frequency of hits in the matched sample was 19.6\%, substantially higher than the 1.0\% observed in the unmatched sample, where each \LLM{} submission was randomly paired with another submission from ICLR 2024. 
    Additionally, we analyzed the most common keywords in the \LLM{} submissions, calculating the frequency of these words in both the matched and unmatched submissions (see Figure \ref{fig:keywords_balance}).

    \item \textit{{Embeddings Similarity}} 
    For each sample, we measure the similarity between the abstracts of matched pairs of submissions using the BERTScores~\cite{zhang2019bertscore}. 
    BERTScore is a metric used to measure textual similarity sensitive to semantic content. This is done by calculating the cosine similarity of BERT embeddings between corresponding tokens in two texts.  
    The average BERTScore for pairs in the matched sample was 0.836, compared to 0.822 in the unmatched sample, confirming the robustness of our content-based matching process ($p$<0.001 in a K-S test).
\end{itemize}

\xhdr{Regression Analysis}\label{si:accept:regession-analysis}
We report the regression coefficients and model summary for the regressions shown in the main paper in Table~\ref{tab:logit_results} and Table~\ref{tab:ols_result}.

\xhdr{Sensitivity Analysis} \label{si:sensitivity}
Our results rely on the assumption that there are no confounders that affect both the probability of receiving an \LLM{} review and the acceptance/rejection outcome. 
Sensitivity analysis is a way of quantifying how the results of our study would change if this assumption is violated \cite{rosenbaum2005sensitivity}.
This notion is quantified by the sensitivity $\Gamma$, which specifies the ratio by which the probability of receiving an \LLM{} review of two matched submissions would need to differ to result in a $p$-value above the significance threshold. Large values of $\Gamma$ correspond to more robust conclusions. 
For the chosen $p=0.05$, we measured the effect of \LLM{} reviews on acceptance decisions. For borderline papers (in the $[5,6)$ bin)
we obtain a $\Gamma$ of 1.07, which implies that, within matched pairs, a submission's probability of receiving an \LLM{} review could take on any value between $1/(1+\Gamma)=0.48$ and $\Gamma/(1+\Gamma)=0.52$ without changing our decision of rejecting the null hypothesis of no effect.

\vspace*{\fill}

\clearpage

\input{tables/acceptance_linear}
\input{tables/acceptance_logistic}
\input{tables/acceptance_stratified_linear_regression}

\clearpage
\vspace*{\fill}

\input{tables/acceptance_stratified_logistic_regression_table}

\vspace*{\fill}

\clearpage

\vspace*{\fill}

\section{Prevalence Checks} 
\label{si:prevalence}

\xhdr{Interpreting GPTZero Predictions Using Word Frequencies} \label{si:prevalence:gptz-interpret}
To enhance the interpretability of GPTZero's prediction, we investigate if reviews containing words that are typically associated with AI-generated text are more likely to be predicted as \LLM{}. To do so, we define a dictionary of all words in our dataset and filter it down to either nouns, verbs, or adjectives. Then, for each word $w$, we compute the probability of a review, $r$, being classified as \LLM{}, $\text{LLM}_r$, if that word is present, denoted as $Pr(\text{LLM}_r|w)$. We illustrate the 30 most predictive words in Table~\ref{tab:word_probabilities} alongside their probability.

\begin{table}[ht]
\normalsize
\centering
\begin{tabular}{rrrr}
\toprule
\multicolumn{4}{c}{Words and Probabilities} \\
\midrule
underscores (0.780), & necessitating (0.747), & delves (0.741), & adaptability (0.731), \\
delved (0.727), & delve (0.722) & elucidated (0.709), & underscore (0.695),  \\
credibility (0.688), & advancements (0.687), & elucidation (0.686), & underpinnings (0.681).\\
equitable (0.679), & perplexing (0.676), & excels (0.674), & intricacies (0.672),\\
persuasiveness (0.670), & delineation (0.667), & elucidate (0.667), & provision (0.658),\\
bolster (0.654), & discourse (0.652), & meticulous (0.652), & endeavors (0.650),\\
tangible (0.650), & commendable (0.645), & showcasing (0.643), & imperative (0.642),\\
encompassing (0.638), & offering (0.633) \\
\bottomrule
\end{tabular}
\caption{\textbf{Predominant LLM-Associated Words}. This table displays the most predictive words in GPTZero classifications. The probability indicates the likelihood that a review containing a specific word is classified as \LLM{}.}
\label{tab:word_probabilities}
\end{table}

 \label{si:prevalence:llm-assisted-thresholds}

\vspace*{\fill}

\clearpage

\vspace*{\fill}

\section{Overlap of Conference Reviewers and Area Chairs} \label{si:overlap}
\noindent We collected data on Area-Chair and reviewer volunteers of several top AI conferences directly from their official web pages. Specifically, we collected data from the conference on ``Artificial Intelligence and Statistics'' (AISTATS), 
the ``International Conference on Learning Representations'' (ICLR), the ``International Conference on Machine Learning'' (ICML), and the ``Conference on Neural Information Processing Systems'' (NeurIPS). We measure the overlap in exact names across the different conferences by taking the Jaccard similarity. In Table~\ref{tab:reviewer_overlap} we illustrate the overlap in reviewers, whereas in Table~\ref{tab:ac_overlap} we show the overlap in area chairs and meta reviewers. 

\begin{table}[ht]
        \centering 
        \begin{tabular}{p{2.25cm}p{1.5cm}p{1.5cm}p{1.5cm}p{1.5cm}p{1.5cm}p{1.5cm}p{1.5cm}p{1.5cm}p{1.5cm}}
    \toprule
    & AISTATS 2023 & AISTATS 2024 & ICML 2022 & ICML 2023 & NeurIPS 2022 & NeurIPS 2023 & ICLR 2023 \\
    \midrule
    AISTATS 2023 & -- & -- & -- & -- & -- & -- & -- \\
    AISTATS 2024 & 0.264 & -- & -- & -- & -- & -- & -- \\
    ICML 2022 & 0.092 & 0.075 & -- & -- & -- & -- & -- \\
    ICML 2023 & 0.095 & 0.082 & 0.327 & -- & -- & -- & -- \\
    NeurIPS 2022 & 0.078 & 0.066 & 0.402 & 0.460 & -- & -- & -- \\
    NeurIPS 2023 & 0.076 & 0.069 & 0.284 & 0.327 & 0.400 & -- & -- \\
    ICLR 2023 & 0.070 & 0.059 & 0.269 & 0.299 & 0.342 & 0.237 & -- \\
    \textbf{ICLR 2024} & \textbf{0.073} & \textbf{0.076} & \textbf{0.261} & \textbf{0.312} & \textbf{0.330} & \textbf{0.534} & \textbf{0.265} \\
    \bottomrule
    \end{tabular}
    \caption{\textbf{Jaccard similarity between reviewers across conferences.} A similarity of 1 indicates full overlap, and a similarity of 0 indicates no reviewer overlap.}
    \label{tab:reviewer_overlap}
\end{table}

\begin{table}[ht]
        \centering 
        \begin{tabular}{p{2.25cm}p{1.5cm}p{1.5cm}p{1.5cm}p{1.5cm}p{1.5cm}p{1.5cm}p{1.5cm}p{1.5cm}p{1.5cm}}
        \toprule
     & AISTATS 2023 & AISTATS 2024 & ICML 2022 & ICML 2023 & NeurIPS 2022 & NeurIPS 2023 & ICLR 2023 \\
    \midrule
    AISTATS 2023 & -- & -- & -- & -- & -- & -- & -- \\
    AISTATS 2024 & 0.528 & -- & -- & -- & -- & -- & -- \\
    ICML 2022 & 0.049 & 0.051 & -- & -- & -- & -- & -- \\
    ICML 2023 & 0.047 & 0.048 & 0.241 & -- & -- & -- & -- \\
    NeurIPS 2022 & 0.043 & 0.036 & 0.237 & 0.652 & -- & -- & -- \\
    NeurIPS 2023 & 0.036 & 0.034 & 0.136 & 0.286 & 0.318 & -- & -- \\
    ICLR 2023 & 0.039 & 0.028 & 0.163 & 0.233 & 0.255 & 0.194 & -- \\
    \textbf{ICLR 2024} & \textbf{0.046} & \textbf{0.054} & \textbf{0.153} & \textbf{0.248} & \textbf{0.242} & \textbf{0.254} & \textbf{0.332} \\
    \bottomrule
    \end{tabular}

    \caption{\textbf{Jaccard similarity between area chairs and meta reviewers across conferences.} A similarity of 1 indicates full overlap, and a similarity of 0 indicates no reviewer overlap.}
    \label{tab:ac_overlap}
\end{table}
\vspace*{\fill}

\vspace*{\fill}

\section{Examples of Matched Submissions}
\label{sec:si:examples}

In Table \ref{table:abstracts}, we provide examples of matched papers abstracts that received an \LLM{} review with those papers that received only human reviews. These examples are randomly sampled from the set of matches found using the procedure described in \MatMeth.

\begin{longtable}{|p{\dimexpr 0.5\textwidth-2\tabcolsep-1.5\arrayrulewidth}|p{\dimexpr 0.5\textwidth-2\tabcolsep-1.5\arrayrulewidth}|}
\caption{Examples of papers' abstracts that received an \LLM{} review  and of abstract of papers that received only human reviews. } \label{table:abstracts} \\
\hline
\textbf{Abstract with \LLM{} reviews} & \textbf{Abstract with human reviews} \\
\hline
\endfirsthead

\multicolumn{2}{c}%
{{\bfseries \tablename\ \thetable{} -- continued from previous page}} \\
\hline
\textbf{AI-assisted Abstract} & \textbf{Human Abstract} \\
\hline
\endhead

\hline \multicolumn{2}{|r|}{{Continued on next page}} \\ \hline
\endfoot

\hline
\endlastfoot

% Your actual content starts here, repeated as per your original table
In this paper, we consider offline reinforcement learning (RL) problems. Within this setting, posterior sampling has been rarely used, perhaps partly due to its explorative nature. The only work using posterior sampling for offline RL that we are aware of is the model-based posterior sampling of Ueara et al.. However, this framework does not permit any tractable algorithm (not even in the linear models) where simulations of posterior samples become challenging, especially in high dimensions. In addition, the algorithm only admits a weak form of guarantees -- Bayesian sub-optimality bounds which depend on the prior distribution. To address these problems, we propose and analyze the use of Markov Chain Monte Carlo methods for offline RL. We show that for low-rank Markov decision processes (MDPs), using the Langevin Monte Carlo (LMC) algorithm, our algorithm obtains the (frequentist) sub-optimality bound that competes against any comparator policy $\pi$ and interpolates between $\tilde{\mathcal{O}}(H^2 d \sqrt{C_{\pi}/ K})$ and $\tilde{\mathcal{O}}(H^2  \sqrt{d C_{\pi}/ K})$, where $C_{\pi}$ is the concentrability coefficient of $\pi$, $d$ is the dimension of the linear feature, $H$ is the episode length, and $K$ is the number of episodes in the offline data. For general MDPs with overparameterized neural network function approximation, we show that our LMC-based algorithm obtains the sub-optimality bounds of $\tilde{\mathcal{O}}(H^{2.5}  \tilde{d}  \sqrt{C_{\pi} /K})$,  where $\tilde{d}$ is the effective dimension of the neural network. Finally, we collaborate our findings with numerical evaluations to demonstrate that LMC-based algorithms could be both efficient and competitive for offline RL in high dimensions. &
Branch-and-bound (B\&B) has long been favored for tackling complex Mixed Integer Programming (MIP) problems, where the choice of branching strategy plays a pivotal role. Recently, Imitation Learning (IL)-based policies have emerged as potent alternatives to traditional rule-based approaches. However, it is nontrivial of acquiring high-quality training samples, and IL often converges to suboptimal variable choices for branching, restricting the overall performance. In response to these challenges, we propose a novel hybrid online and offline reinforcement learning (RL) approach to enhance the branching policy by cost-effective training sample augmentation. In online phase, we train an online RL agent to dynamically decide the sample generation processes, drawing from either the learning-based policy or the expert policy. The objective here is to strike an optimal balance between the exploration and exploitation of the sample generation process. In offline phase, a value function is trained to fit the cumulative reward for each decision and to filter the samples with high cumulative returns. This dual-purpose function not only reduces training complexity but also enhances the quality of the samples. To assess the efficacy of our proposed data augmentation mechanism, we conduct comprehensive evaluations across a range of MIP problems. The results consistently show that our method excels in making superior branching decisions compared to state-of-the-art learning-based models and the open-source solver SCIP. Notably, it even often outperforms the commercial solver Gurobi. \\ \hline

In this paper, we investigate the generalization error of deep physical models with latent variables. Deep physical models, such as Hamiltonian Neural Networks, are neural network models for learning equations of motion from observational data of physical phenomena and have attracted much attention in recent years. In particular, in such cases, the data are not completely random, but rather given as random trajectories. We provide an error bound for the case where the training data are given in such a way. Our results show that it is important to collect data from many trajectories, rather than simply collecting a large number of data, to improve generalization performance. In addition, an important application of the combination of deep physics models with latent variables is the interpolation of images from videos while preserving the laws of physics, such as the energy conservation law. However, when the frame interval of the video is large, it can be difficult to preserve the laws of physics. In this paper, we show that it is possible to interpolate the images from videos so that the laws of physics are preserved, provided that the generalization error is sufficiently small & 

Scientific processes are often modelled by sets of differential equations. As datasets grow, individually fitting these models and quantifying their uncertainties becomes a computationally challenging task. In this paper, we focus on improving the scalability of a particular class of stochastic dynamical model, called latent force models. These offer a balance between data-driven and mechanistic inference in dynamical systems, achieved by deriving a kernel function over a low-dimensional latent force. However, exact computation of posterior kernel terms is rarely tractable, requiring approximations for complex scenarios such as nonlinear dynamics. We overcome this issue by posing the problem as meta-learning the class of latent force models corresponding to a set of differential equations. By employing a deep kernel along with a sensible function embedding, we demonstrate the ability to extrapolate from simulations to real experimental datasets. Finally, we show how our model scales compared with other approximations. \\ \hline

Image restoration problems are typically ill-posed in the sense that each degraded image can be restored in infinitely many valid ways. To accommodate this, many works generate a diverse set of outputs by attempting to randomly sample from the posterior distribution of natural images given the degraded input. Here we argue that this strategy is commonly of limited practical value because of the heavy tail of the posterior distribution. Consider for example inpainting a missing region of the sky in an image. Since there is a high probability that the missing region contains no object but clouds, any set of samples from the posterior would be entirely dominated by (practically identical) completions of sky. However, arguably, presenting users with only one clear sky completion, along with several alternative solutions such as airships, birds, and balloons, would better outline the set of possibilities. In this paper, we initiate the study of **meaningfully diverse** image restoration. We explore several post-processing approaches that can be combined with any diverse image restoration method to yield semantically meaningful diversity. Moreover, we propose a practical approach for allowing diffusion based image restoration methods to generate meaningfully diverse outputs, while incurring only negligent computational overhead. We conduct extensive user studies to analyze the proposed techniques, and find the strategy of reducing similarity between outputs to be significantly favorable over posterior sampling. &

Given an image of a natural scene, we are able to quickly decompose it into a set of components such as objects, lighting, shadows, and foreground. We can then picture how the image would look if we were to recombine certain components with those from other images, for instance producing a scene with a set of objects from our bedroom and animals from a zoo under the lighting conditions of a forest even if we have never seen such a scene in real life before. We present a method to decompose an image into such compositional components. Our approach, Decomp Diffusion, is an unsupervised method which, when given a single image, infers a set of different components in the image, each represented by a diffusion model. We demonstrate how components can capture different factors of the scene, ranging from global scene descriptors (shadows, foreground, facial expression) to local scene descriptors (objects). We further illustrate how inferred factors can be flexibly composed, even with factors inferred from other models, to generate a variety of scenes sharply different than those seen in training time. \\ \hline

The excessive computational requirements of modern artificial neural networks (ANNs) are posing limitations on the machines that can run them. Sparsification of ANNs is often motivated by time, memory and energy savings only during model inference, yielding no benefits during training. A growing body of work is now focusing on providing the benefits of model sparsification also during training. While these methods greatly improve the training efficiency, the training algorithms yielding the most accurate models still materialize the dense weights, or compute dense gradients during training. We propose an efficient, always-sparse training algorithm which improves the accuracy over previous methods. Additionally, our method has excellent scaling to larger and sparser models, supported by its linear time complexity with respect to the model width during training and inference. We evaluate our method on CIFAR-10/100 and ImageNet using ResNet, VGG, and ViT models, and compare it against a range of sparsification methods. &

Deep neural networks have demonstrated remarkable performance in various tasks. With a growing need for sparse deep learning, model compression techniques, especially pruning, have gained significant attention. However, conventional pruning techniques can inadvertently exacerbate algorithmic bias, resulting in unequal predictions. To address this, we define a fair pruning task where a sparse model is derived subject to fairness requirements. In particular, we propose a framework to jointly optimize the pruning mask and weight update processes with fairness constraints. This framework is engineered to compress models that maintain performance while ensuring fairness in a single execution. To this end, we formulate the fair pruning problem as a novel constrained bi-level optimization task and derive efficient and effective solving strategies. We design experiments spanning various datasets and settings to validate our proposed method. Our empirical analysis contrasts our framework with several mainstream pruning strategies, emphasizing our method's superiority in maintaining model fairness, performance, and efficiency. \\ \hline

Single Domain Generalization Object Detection (S-DGOD) is a challenging yet practical task, where we only have access to data from one specific source domain to train an object detection network, but have to generalize to numerous unseen target domains. Recent works point out that the learning dynamics of Deep Neural Networks (DNNs) are biased by gradient descent to learn simple semantics, which are usually non-causal and spuriously correlated to the ground truth labels, as a result, DNN-based object detection networks fail to consistently generalize well in the Out-of-Domain (OoD) scenario. In this paper, we focus on S-DGOD based on theoretical analysis, exploring a classic and widely-used approach, Generalizable Reweighting (GRW), which iteratively reweightes the training samples to improve generalization performance. In our theoretical analysis, we first identify that the vanilla GRW hardly outperforms Empirical Risk Minimization (ERM) in the S-DGOD scenario. To provide a generalization guarantee, we further derive Certifiable Feature Perturbation (CFP) based on our theory, which aims to train a robust object detection network against additional perturbations added to the extracted features. We demonstrate that GRW works well with CFP in achieving OoD generalization, thus, surpassing ERM by a large margin under worse conditions. This brand new reweighting strategy is named Certifiable Reweighting (CARD). Our extensive experiments show that the proposed CARD achieves SOTA performance compared to baseline methods on the five urban-scene S-DGOD benchmarks. &

Few-shot object detection (FSOD) benchmarks have advanced techniques for detecting new categories using limited annotations. Existing FSOD benchmarks re-purpose well-established datasets like COCO by partitioning categories into base and novel classes for pre-training and fine-tuning respectively. However, these benchmarks do not reflect how FSOD is deployed in practice. Rather than pre-training on only a small number of categories, we argue that it is more practical to download a foundational model (e.g., a vision-language model (VLM) pretrained on web-scale data) and finetune it for specific applications. Surprisingly, we find that zero-shot inference from foundational VLMs like GroundingDINO significantly outperform state-of-the-art methods (48.3 vs. 33.1 AP) on COCO, suggesting that few-shot detection should be reframed in the context of foundation models. In this work, we propose a new FSOD benchmark protocol that evaluates detectors pre-trained on any external dataset (not including the target dataset), and finetuned on K-shot annotations per C target classes. Further, we note that FSOD benchmarks are actually federated datasets, which are exhaustively annotated for a single category only on a subset of data. We leverage this insight and propose simple strategies for fine-tuning VLMs to improve FSOD. We demonstrate the effectiveness of our approach on LVIS and nuImages \\ \hline

Synthetic data has the distinct advantage of building a large-scale labeled dataset for almost free. Still, it should be carefully integrated into learning; otherwise, the expected performance gains are difficult to achieve. The biggest hurdle for synthetic data to achieve increased training performance is the domain gap with the (real) test data. As a common solution to deal with the domain gap, the sim2real transformation is used, and its quality is affected by three factors: i) the real data serving as a reference when calculating the domain gap, ii) the synthetic data chosen to avoid the transformation quality degradation, and iii) the synthetic data pool from which the synthetic data is selected. In this paper, we investigate the impact of these factors on maximizing the effectiveness of synthetic data in training in terms of improving learning performance and acquiring domain generalization ability--two main benefits expected of using synthetic data. As an evaluation metric for the second benefit, we introduce a method for measuring the distribution gap between two datasets, which is derived as the normalized sum of the Mahalanobis distances of all test data. As a result, we have discovered several important findings that have never been investigated or have been used previously without accurate understanding. We expect that these findings can break the current trend of either naively using or being hesitant to use synthetic data in machine learning due to the lack of understanding, leading to more appropriate use in future research. &

The performance of machine learning models on new data is critical for their success in real-world applications. However, the model's performance may deteriorate if the new data is sampled from a different distribution than the training data. Current methods to detect shifts in the input or output data distributions have limitations in identifying model behavior changes. In this paper, we define explanation shift as the statistical comparison between how predictions from training data are explained and how predictions on new data are explained. We propose explanation shift as a key indicator to investigate the interaction between distribution shifts and learned models.  We introduce an Explanation Shift Detector that operates on the explanation distributions, providing more sensitive and explainable changes in interactions between distribution shifts and learned models. We compare explanation shifts with other methods that are based on distribution shifts, showing that monitoring for explanation shifts results in more sensitive indicators for varying model behavior. We provide theoretical and experimental evidence and demonstrate the effectiveness of our approach on synthetic and real data. Additionally, we release an open-source Python package, skshift, which implements our method and provides usage tutorials for further reproducibility. \\ \hline

Irregular sampling intervals and missing values in real-world time series data present challenges for conventional methods that assume consistent intervals and complete data. Neural Ordinary Differential Equations (Neural ODEs) offer an alternative approach, utilizing neural networks combined with ODE solvers to learn continuous latent representations through parameterized vector fields. Neural Stochastic Differential Equations (Neural SDEs) extend Neural ODEs by incorporating a diffusion term, although this addition is not trivial, particularly when addressing irregular intervals and missing values. Consequently, careful design of drift and diffusion functions is crucial for maintaining stability and enhancing performance, while incautious choices can result in adverse properties such as the absence of strong solutions, stochastic destabilization, or unstable Euler discretizations, significantly affecting Neural SDEs' performance. In this study, we propose three stable classes of Neural SDEs: Langevin-type SDE, Linear Noise SDE, and Geometric SDE. Then, we rigorously demonstrate their robustness in maintaining excellent performance under distribution shift, while effectively preventing overfitting. To assess the effectiveness of our approach, we conduct extensive experiments on four benchmark datasets for interpolation, forecasting, and classification tasks, and analyze the robustness of our methods with 30 public datasets under different missing rates. Our results demonstrate the efficacy of the proposed method in handling real-world irregular time series data. &

Limited data availability poses a major obstacle in training deep learning models for financial applications. Synthesizing financial time series to augment real-world data is challenging due to the irregular and scale-invariant patterns uniquely associated with financial time series - temporal dynamics that repeat with varying duration and magnitude. Such dynamics cannot be captured by existing approaches, which often assume regularity and uniformity in the underlying data. We develop a novel generative framework called FTS-Diffusion to model irregular and scale-invariant patterns that consists of three modules. First, we develop a scale-invariant pattern recognition algorithm to extract recurring patterns that vary in duration and magnitude. Second, we construct a diffusion-based generative network to synthesize segments of patterns. Third, we model the temporal transition of patterns in order to aggregate the generated segments. Extensive experiments show that FTS-Diffusion generates synthetic financial time series highly resembling observed data, outperforming state-of-the-art alternatives. Two downstream experiments demonstrate that augmenting real-world data with synthetic data generated by FTS-Diffusion reduces the error of stock market prediction by up to 17.9\%. To the best of our knowledge, this is the first work on generating intricate time series with irregular and scale-invariant patterns, addressing data limitation issues in finance. \\ \hline

Surrogate neural network-based partial differential equation (PDE) solvers have the potential to solve PDEs in an accelerated manner, but they are largely limited to systems featuring predetermined problem sizes or fixed PDE parameters. We propose Specialized Neural Accelerator-Powered Domain Decomposition Methods (SNAP-DDM), a DDM-based approach to PDE solving in which subdomain problems containing arbitrary boundary conditions and geometric parameters are accurately solved using an ensemble of specialized neural operators.  We tailor SNAP-DDM to 2D electromagnetics and fluidic flow problems and show how innovations in network architecture and loss function engineering can produce specialized surrogate subdomain solvers with near unity accuracy.  We also show how these solvers can be used with standard DDM algorithms to accurately solve freeform electromagnetics and fluids problems with a wide range of domain sizes. &

Recent work provides promising evidence that Physics-Informed Neural Networks (PINN) can efficiently solve partial differential equations (PDE). However, previous works have failed to provide guarantees on the worst-case residual error of a PINN across the spatio-temporal domain - a measure akin to the tolerance of numerical solvers - focusing instead on point-wise comparisons between their solution and the ones obtained by a solver on a set of inputs. In real-world applications, one cannot consider tests on a finite set of points to be sufficient grounds for deployment, as the performance could be substantially worse on a different set. To alleviate this issue, we establish tolerance-based correctness conditions for PINNs over the entire input domain. To verify the extent to which they hold, we introduce partial-CROWN: a general, efficient and scalable post-training framework to bound PINN residual errors. We demonstrate its effectiveness in obtaining tight certificates by applying it to two classically studied PDEs - Burgers' and Schrödinger's equations -, and two more challenging ones with real-world applications - the Allan-Cahn and Diffusion-Sorption equations. \\ \hline

Information retrieval (IR) plays a crucial role in locating relevant resources from vast amounts of data, and its applications have evolved from traditional knowledge bases to modern retrieval models (RMs). The emergence of large language models (LLMs) has further revolutionized the IR field by enabling users to interact with search systems in natural languages. In this paper, we explore the advantages and disadvantages of LLMs and RMs, highlighting their respective strengths in understanding user-issued queries and retrieving up-to-date information. To leverage the benefits of both paradigms while circumventing their limitations, we propose InteR, a novel framework that facilitates information refinement through synergy between RMs and LLMs. InteR allows RMs to expand knowledge in queries using LLM-generated knowledge collections and enables LLMs to enhance prompt formulation using retrieved documents. This iterative refinement process augments the inputs of RMs and LLMs, leading to more accurate retrieval. Experiments on large-scale retrieval benchmarks involving web search and low-resource retrieval tasks demonstrate that InteR achieves overall superior zero-shot retrieval performance compared to state-of-the-art methods, even those using relevance judgment. &

Large language models (LLMs) are initially pretrained for broad capabilities and then finetuned with instruction-following datasets to improve their performance in interacting with humans. Despite advances in finetuning, a standardized guideline for selecting high-quality datasets to optimize this process remains elusive. In this paper, we first propose InstructMining, an innovative method designed for automatically selecting premium instruction-following data for finetuning LLMs. Specifically, InstructMining utilizes natural language indicators as a measure of data quality, applying them to evaluate unseen datasets. During experimentation, we discover that double descent phenomenon exists in large language model finetuning. Based on this observation, we further leverage BlendSearch to help find the best subset among the entire dataset (i.e., 2,532 out of 100,000). Experiment results show that InstructMining-7B achieves state-of-the-art performance on two of the most popular benchmarks: LLM-as-a-judge and OpenLLM benchmark. \\ \hline

We present chain-of-knowledge (CoK), a novel framework that augments large language models (LLMs) by dynamically incorporating grounding information from heterogeneous sources. It results in more factual rationales and reduced hallucination in generation. Specifically, CoK consists of three stages: reasoning preparation, dynamic knowledge adapting, and answer consolidation. Given a knowledge-intensive question, CoK first prepares several preliminary rationales and answers while identifying the relevant knowledge domains.If there is no majority consensus among the answers from samples, CoK corrects the rationales step by step by adapting knowledge from the identified domains.These corrected rationales can plausibly serve as a better foundation for the final answer consolidation.Unlike prior studies that primarily use unstructured data, CoK also leverages structured knowledge sources such as Wikidata and tables that provide more reliable factual information.To access both unstructured and structured knowledge sources in the dynamic knowledge adapting stage, we propose an adaptive query generator that allows the generation of queries for various types of query languages, including SPARQL, SQL, and natural sentences. Moreover, to minimize error propagation between rationales, CoK corrects the rationales progressively using preceding corrected rationales to generate and correct subsequent rationales. Extensive experiments show that CoK consistently improves the performance of LLMs on knowledge-intensive tasks across different domains. &

Retrieval-augmented language models (RALMs) hold promise to produce language understanding systems that are are factual, efficient, and up-to-date. An important desideratum of RALMs, is that retrieved information helps model performance when it is relevant, and does not harm performance when it is not. This is particularly important in multi-hop reasoning scenarios, where misuse of irrelevant evidence can lead to cascading errors. However, recent work has shown that retrieval augmentation can sometimes have a negative effect on performance. In this work, we present a thorough analysis on five open-domain question answering benchmarks, characterizing cases when retrieval reduces accuracy. We then propose two methods to mitigate this issue. First, a simple baseline that filters out retrieved passages that do not entail question-answer pairs according to a natural language inference (NLI) model. This is effective in preventing performance reduction, but at a cost of also discarding relevant passages. Thus, we propose a method for automatically generating data to fine-tune the language model to properly leverage retrieved passages, using a mix of relevant and irrelevant contexts at training time. We empirically show that even 1,000 examples suffice to train the model to be robust to irrelevant contexts while maintaining high performance on examples with relevant ones. \\ \hline

\end{longtable}
\end{document}

%% file: tables/dataset.tex
\begin{table}[h]
\centering
\begin{tabular}{p{1.5cm}p{2.5cm}p{2.5cm}p{2.5cm}p{3cm}}
\toprule
Year & Reviews & Submissions  & Acceptance  & \LLM{} reviews\\
\midrule
2018  & 2921 & 1007    & 36.0\% & 57 \\
2019  & 4734 & 1569  & 31.5\% & 95 \\
2020  & 7783 & 2593  & 26.5\% & 123 \\
2021  & 11488 & 3009 & 29.1\% & 216 \\
2022  & 13161 & 3422 & 32.0\% & 164 \\
2023 & 18575 & 4955 & 24.3\% & 176 \\
2024 & 28028 & 7404  & 30.5\% & 4887 \\ \midrule
Total & 86690 & 23959 & --- & --- \\
\bottomrule
\end{tabular}
\caption{Number of ICLR reviews and submissions per year. We report the number of \LLM{} reviews detected by GPTZero (without correcting for the model's false positive rate).}
\label{tab:summary}
\end{table}

%% file: tables/scores_ordinal_regression.tex
\begin{table}[h]
\centering
\begin{tabular}{p{5cm}ccc}
\toprule
\textbf{Variable} & \textbf{Coefficient} & \textbf{Std.Error} & \textbf{P-value} \\
\midrule
$\gamma$     & 0.1463 & 0.0335 & 1.336e-05 \\
$\alpha_1$   & -2.1794 & 0.2230& < 2.2e16  \\
$\alpha_3$   & 0.7121 & 0.2036 &   0.0004 \\
$\alpha_5$   & 2.1793 & 0.2049 & < 2.2e-16 \\
$\alpha_6$   & 3.9637 & 0.2084 & < 2.2e-16  \\
$\alpha_8$   & 7.6799 & 0.2723 & < 2.2e-16         \\
$\beta_3$    & 0.9800 & 0.2064 & 2.110e-06\\
$\beta_5$    & 1.5785 & 0.2070 & 2.813e-14\\
$\beta_6$    & 2.4094 & 0.2081 & < 2.2e-16 \\
$\beta_8$    & 2.9347 & 0.2143 & < 2.2e-16 \\
\midrule
\textbf{Model Summary:} & & & \\
Number of observations & {19332} \\
Log-likelihood & {-6346.89} \\
AIC & {12705.78} \\
\bottomrule
\end{tabular}
\caption{\textbf{Ordinal regression results.} Coefficients and model summary (see Eq~\ref{eq:rq2})}
\label{tab:regression_results}
\end{table}

%% file: tables/acceptance_linear.tex
\begin{table}[h]
\centering
\begin{tabular}{p{5cm}ccc}
\toprule
\textbf{Variable} &    \textbf{Coefficient} & \textbf{Std.Error} &  \textbf{P-value} \\ 
\midrule
Intercept & -0.5327 & 0.041  & 0.000   \\
$\beta$ & 0.1289 & 0.057  & 0.024  \\
\midrule
\textbf{Model Summary:} &  \\
Number of Observations: & 5180 \\
Df Residuals: & 5178 \\
Psuedo R-squared.: & 0.00073 \\
Log-Likelihood: & -3450.0\\ \bottomrule
\end{tabular}
\caption{\textbf{Logistic regression results.} Coefficients and model summary (see  Eq.~\ref{eq:rq2}).}
\label{tab:logit_results}
\end{table}

%% file: tables/acceptance_logistic.tex
\begin{table}[h]
\centering
\begin{tabular}{p{5cm}ccc}
\toprule
\textbf{Variable} &    \textbf{Coefficient} & \textbf{Std.Error} &  \textbf{P-value} \\ 
\midrule
Intercept & 0.3699 & 0.010  & 0.000   \\
$\beta$ & 0.0305 & 0.014  & 0.024  \\
\midrule
\textbf{Model Summary:} &  \\
Number of Observations: & 5180 \\
Df Residuals: & 5178 \\
Adj. R-squared.: & 0.001 \\
Log-Likelihood: & -3616.6\\ \bottomrule
\end{tabular}
\caption{\textbf{Ordinary least squares.} Coefficients and model summary (see  Eq.~\ref{eq:rq2b}).}
\label{tab:ols_result}
\end{table}

%% file: tables/acceptance_stratified_linear_regression.tex
\begin{table}[h]
\centering
\begin{tabular}{p{5cm}ccc}
\toprule
\textbf{Variable} & \textbf{Coefficient} & \textbf{Std.Err} &  \textbf{P-value} \\
\midrule
Intercept & 0.2143 & 0.049 & 0.000 \\
Bin=2-3 & -0.0589 & 0.053 & 0.267 \\
Bin=3-4 & 0.0481 & 0.052 & 0.270 \\
Bin=4-5 & -0.0569 & 0.051 & 0.347 \\
Bin=5-6 & 0.5218 & 0.051 & 0.000 \\
Bin=6-7 & 0.7554  & 0.070 & 0.000 \\
Bin=7-8 & 0.7857 & 0.412 & 0.057 \\
$\beta \cdot $ Bin=1-2 & -0.0143 & 0.069 & 0.836 \\
$\beta \cdot $ Bin=2-3 & 0.0301 & 0.029 & 0.299 \\
$\beta \cdot $ Bin=3-4 & 0.0426 & 0.023 & 0.069 \\
$\beta \cdot $ Bin=4-5 & 0.0110 & 0.021 & 0.607 \\
$\beta \cdot $ Bin=5-6 & 0.0486 & 0.022 & 0.024 \\
$\beta \cdot $ Bin=6-7 & -0.0152 & 0.071 & 0.832 \\
$\beta \cdot $ Bin=7-8 & 6.37e$^{-15}$ & 0.578 & 1.000 \\
\midrule
\textbf{Model Summary}\\
No. Observations: &{5180} \\
Df Residuals: & {5166} \\
Pseudo R-squ.: & {0.2290} \\
Log-Likelihood: & {-2659.0} \\
\bottomrule
\end{tabular}
\caption{\textbf{Stratified Regression Analysis (Linear Regression).} Estimated coefficients to compute the average increase in the acceptance rate of submissions that received \LLM{} reviews conditioned on the average human score that the submission received. }
\end{table}

%% file: tables/acceptance_stratified_logistic_regression_table.tex
\begin{table}[h]
\centering
\begin{tabular}{p{5cm}ccc}
\toprule
\textbf{Variable} & \textbf{Coefficient} & \textbf{Std.Err} &  \textbf{P-value} \\
\midrule
Intercept & -1.2993 & 0.291 & 0.000 \\
Bin=2-3 & -0.3937 & 0.322 & 0.222 \\
Bin=3-4 & -0.3786 & 0.312 & 0.225 \\
Bin=4-5 & 0.2659 & 0.303 & 0.381 \\
Bin=5-6 & 2.3251 & 0.303 & 0.000 \\
Bin=6-7 & 4.7650  & 0.775 & 0.000 \\
Bin=7-8 & 29.9196 & 1.64e$^{6}$ & 1.000 \\
$\beta\cdot$Bin=1-2 & -0.0870 & 0.417 & 0.835 \\
$\beta\cdot$Bin=2-3 & 0.2132 & 0.189 & 0.259 \\
$\beta\cdot$Bin=3-4 & 0.2916 & 0.150 & 0.052 \\
$\beta\cdot$Bin=4-5 & 0.0563 & 0.119 & 0.635 \\
$\beta\cdot$Bin=5-6 & 0.2675 & 0.124 & 0.031 \\
$\beta\cdot$Bin=6-7 & -0.4212 & 0.930 & 0.0.651 \\
$\beta\cdot$Bin=7-8 & -11.9045 & 1.64e$^{6}$ & 1.000 \\
\midrule
\textbf{Model Summary}\\
No. Observations: &{5180} \\
Df Residuals: & {5166} \\
Pseudo R-squ.: & {0.2290} \\
Log-Likelihood: & {-2659.0} \\
\bottomrule
\end{tabular}
\caption{\textbf{Stratified Regression Analysis (Logistic Regression).} Estimated coefficients to compute the increase in the log-odds acceptance rate of submissions that received \LLM{} reviews conditioned on the average human score that the submission received. }
\end{table}